\documentclass[11pt,a4paper]{article}
\usepackage[utf8]{inputenc}

\usepackage{jheppub}

\usepackage{amsfonts}
\usepackage{amsmath}
\usepackage{amssymb}
\usepackage{amsthm}
\usepackage{bm}
\usepackage{subcaption}
\usepackage[scr=boondoxo, bb=pazo]{mathalpha}
\usepackage{extarrows}
\usepackage{float}
\usepackage{graphicx}
\usepackage{mathtools}
\usepackage{physics}
\usepackage{verbatim}
\usepackage{wrapfig}

\usepackage{hyperref}

\def\tilde{\widetilde}
\def\t{\tilde}

\def\bar{\overline}
\def\b{\bar}

\newcommand{\N}{{\mathbb N}}

\def\1{{\mathbb 1}}

\title{Null states from large superpositions of two-sided black holes}

\author[]{John Gardiner}

\affiliation[]{Mani L. Bhaumik Institute for Theoretical Physics \\
	Department of Physics and Astronomy, University of California, Los Angeles, CA 90095, USA }

\emailAdd{johngardiner@ucla.edu}

\abstract{
	A standard insight of the AdS/CFT correspondence is that some aspects of the geometry of a bulk state are encoded in the entanglement structure of its dual boundary state.
	As entanglement is not a linear quantum observable, this means that geometry in a quantum theory of gravity should likewise not be a linear observable.
	This allows for linear dependencies between states with distinct geometries.
	We explore linear dependencies between certain states with simple geometric duals: states made up of $n$ copies of a thermofield double state and the states obtained from this one by permuting the $n$ right hand sides.
	There are $n!$ such states, all dual to distinct geometries.
	We derive expressions for the maximum fidelity between one such state and a linear combination of the others, and see that this fidelity approaches 1 as the number $n$ of black holes increases.
	We also consider the possibility of obtaining a single thermofield double state as the partial trace of a superposition of states whose topology does not connect the two untraced sides.
	We derive lower bounds for the fidelity between the thermofield double state and such partial traces and comment on the conceptual implications of the existence of such states.
}

\begin{document}
\maketitle

\section{Introduction}
A standard insight from holography is that some aspects of the geometry of spacetime are encoded in the entanglement of the dual holographic boundary state \cite{VanRaamsdonk:2016exw}.
Entanglement is nonlinear, i.e.\ the entanglement of a superposition can be different from the entanglement of the terms in the superposition, so it is not a quantum observable.
This leads to the observation that there cannot be a general quantum observable corresponding to geometry \cite{Berenstein:2016pcx}, except perhaps approximately \cite{Almheiri:2016blp}.
This fact is made manifest in situations where a geometric state is equal to the superposition of states with different geometries \cite{Berenstein:2017abm, VanRaamsdonk:2010pw}.

Here we provide an elementary example of such a situation.
The geometries in question all consist of some number of disjoint copies of a two-sided black hole, and are distinguished from each other only by how the several left and right black hole exteriors are joined together by shared black hole interiors.
With $n$ copies of a two-sided black hole, there are $n!$ states obtained by permuting the exteriors on one side.\footnotemark
\footnotetext{Such a permutation includes a permutation of the boundaries and is thus, of course, a nontrivial diffeomorphism and not merely a gauge redundancy.}
We argue on general grounds that, for a sufficiently large number $n$ of copies, these states are not linearly independent, but can be superposed to give a null state (or at least a state that is ``approximately null" in a sense we make precise).
The existence of this linear dependency implies that we can write a geometry of a number of copies of a two-sided black hole as a superposition of states with the exteriors joined up differently.

On its face, the possibility of writing a configuration of wormholes as a superposition of wormholes connected up differently raises some conceptual questions.
For example, if observers jump into opposite sides of one of $n$ AdS-Schwarzschild black holes, they could in principle meet each other in the black hole interior.
Yet our goal is to rewrite such a state of $n$ black holes as a superposition of different geometries in most of which meeting is impossible because the observers' black holes are distinct and disconnected.
\begin{figure}[t]	
	\centering
	\begin{subfigure}[t]{0.45\linewidth}
		\centering
		\includegraphics[width=\linewidth]{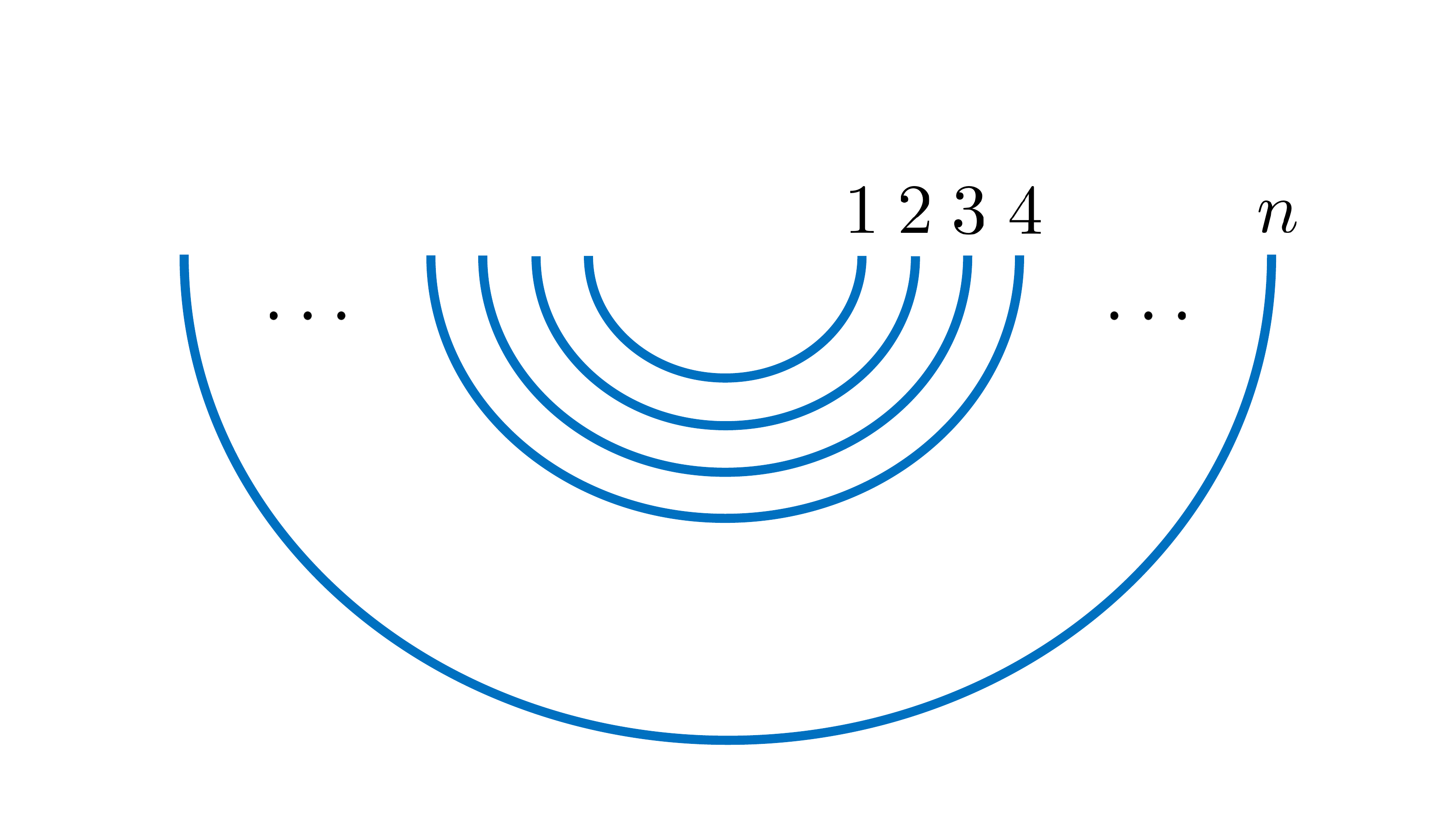}
		\caption{$n$ thermofield double states}\label{fig:TFDstates_unpermuted}		
	\end{subfigure}
	\quad
	\begin{subfigure}[t]{0.45\linewidth}
		\centering
		\includegraphics[width=\linewidth]{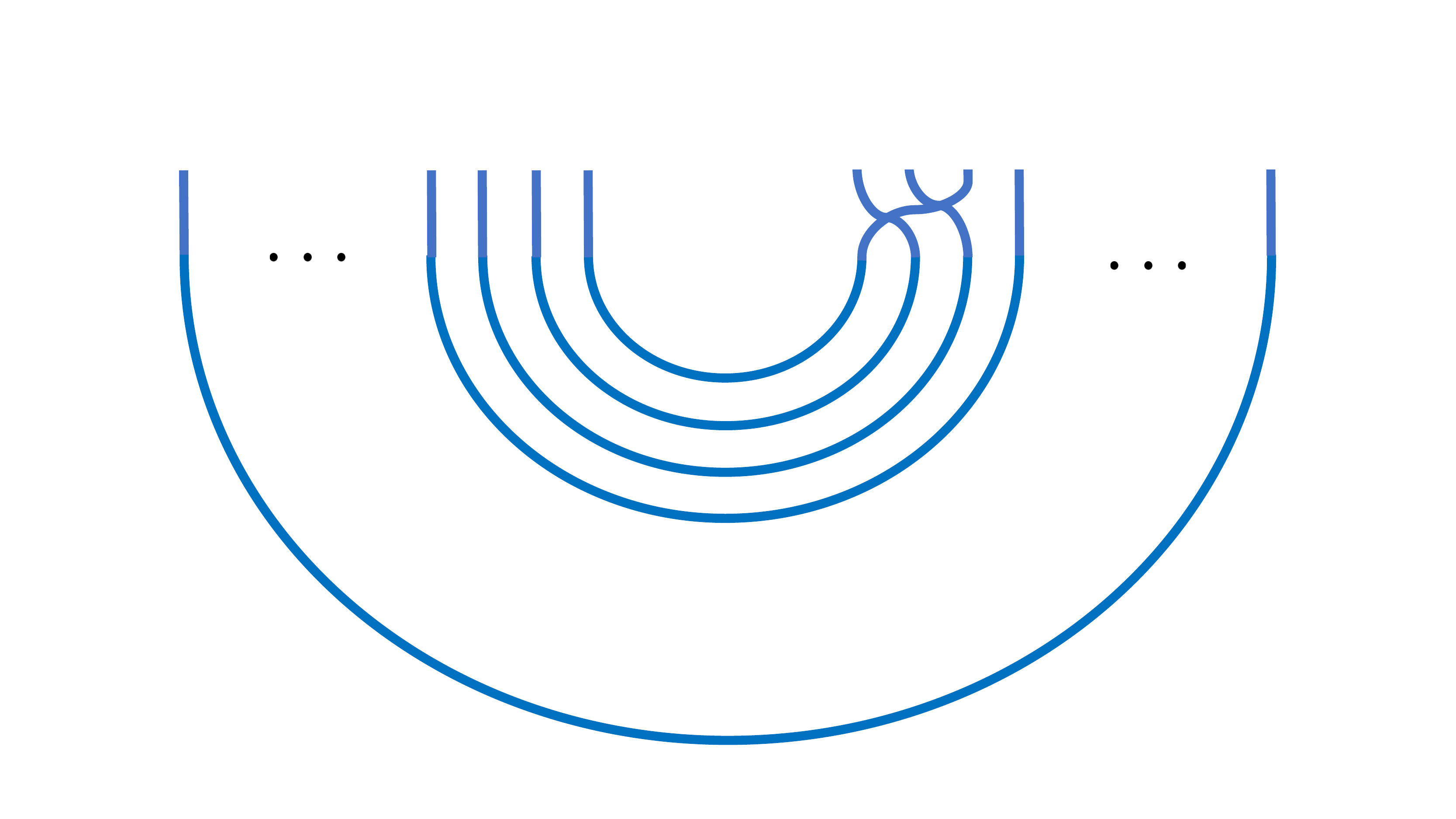}
		\caption{
			TFD states permuted by $(1\ 3\ 2)$
		}\label{fig:TFDstate_permuted}
	\end{subfigure}
	\caption{
		Two states:
		(a) A schematic diagram of the Euclidean path integral preparing $n$ copies of a thermofield double state.
		(b) Diagram of $n$ thermofield double states permuted by $(1\ 3\ 2)$.
		We denote this state by $\ket{(1\ 3\ 2)}$.
	}\label{fig:TFDstates}
\end{figure}

We also explore another possibility that makes the conceptual question sharper.
Is it possible for a superposition of geometries, none of which connect the two observers, to be equivalent to a connected two-sided black hole geometry?
More precisely, can we write a single two-sided black hole as a partial trace of a superposition of geometries that are all disconnected?
In this case, were the observers to jump in, they would meet in the middle, even though this is impossible in any term in the superposition considered separately.
This would imply that there is no quantum observable corresponding to the question of whether the two observers meet.
We argue that such superpositions do exist approximately in the limit of large number $n$ of black holes.
\begin{figure}[t]
	\centering
	\includegraphics[width=\linewidth]{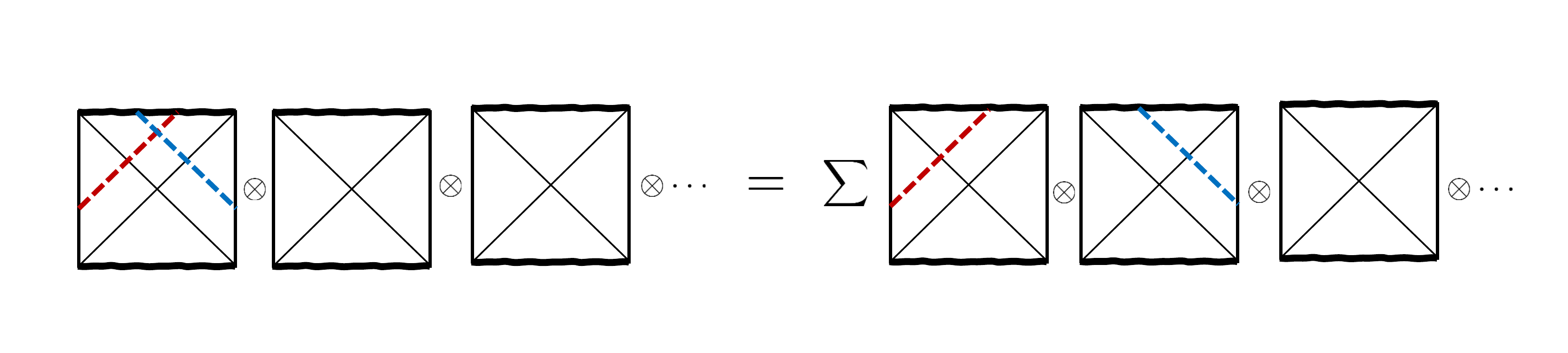}
	\caption{
		Left: Alice (red) and Bob (blue) jump into a black hole. Right: their trajectories do not intersect in any term of an equivalent superposition.
	}\label{fig:Alice_and_Bob}
\end{figure}

\section{Some illustrative examples}\label{example}
We illustrate the idea with a very simple example involving qubits shared between two parties, Alice and Bob.
Suppose Alice and Bob share 3 different Bell pairs of the form $\ket{0}\ket{0}+\ket{1}\ket{1}$.
Suppose further that after Alice and Bob divvy up the qubits, Bob loses track of which of his qubits are which.
In other words, some unknown permutation of Bob's three qubits occurs.
There are 6 possibilities for the state shared by Alice and Bob.
Explicitly, we have
\[
\ket{(1)} = \frac{1}{\sqrt{8}}\sum_{ijk}\ket{i}_A\ket{j}_A\ket{k}_A\otimes\ket{i}_B\ket{j}_B\ket{k}_B,
\]
\[
\ket{(23)} = \frac{1}{\sqrt{8}}\sum_{ijk}\ket{i}_A\ket{j}_A\ket{k}_A\otimes\ket{i}_B\ket{k}_B\ket{j}_B,
\]
\[
\ket{(12)} = \frac{1}{\sqrt{8}}\sum_{ijk}\ket{i}_A\ket{j}_A\ket{k}_A\otimes\ket{j}_B\ket{i}_B\ket{k}_B,
\]
\[
\ket{(13)} = \frac{1}{\sqrt{8}}\sum_{ijk}\ket{i}_A\ket{j}_A\ket{k}_A\otimes\ket{j}_B\ket{k}_B\ket{i}_B,
\]
\[
\ket{(123)} = \frac{1}{\sqrt{8}}\sum_{ijk}\ket{i}_A\ket{j}_A\ket{k}_A\otimes\ket{k}_B\ket{i}_B\ket{j}_B,
\]
\[
\ket{(132)} = \frac{1}{\sqrt{8}}\sum_{ijk}\ket{i}_A\ket{j}_A\ket{k}_A\otimes\ket{k}_B\ket{j}_B\ket{i}_B,
\]
labeled here by permutations.
The situation is that they no longer know which of Alice's qubits are entangled with which of Bob's.
To what extent are Alice and Bob able to determine which qubits are entangled with which?
The task cannot be completed deterministically, as there are nontrivial overlaps between the six states.
For example, $\braket{(1)}{(12)} = \frac{1}{2}$ and $\braket{(1)}{(123)} = \frac{1}{4}$.
An immediate consequence of this is that ``which qubit is entangled with which" is not a quantum observable.

The situation is improved if Alice and Bob share entangled qudit pairs $\sum_{i=1}^N\frac{1}{\sqrt{N}}\ket{i}_A\ket{i}_B$ where $N$ is large.
Then the nontrivial overlaps between states are order $\mathcal{O}(N^{-1})$ or smaller, making the different situations approximately distinguishable.
Alice and Bob can engineer joint measurements that will correctly determine which qudits are entangled with high probability.
This is true even when Alice and Bob share large numbers of qudit pairs.
In effect, more entanglement between the qudits helps ``keep track" of which are matched with which.

An interesting situation arises when Alice and Bob share sufficiently large numbers of entangled pairs.
Let $n$ be the number of shared pairs.
Then there are $n!$ states labeled by permutations, which live in a joint Hilbert space of size $N^{2n}$.
As $n!$ grows faster than $N^{2n}$, for sufficiently large $n$ the $n!$ states labeled by permutations must be linearly dependent.
In fact, linear dependence happens much sooner than the above dimension-counting argument suggests.
We get a nontrivial linear dependence whenever $n > N$.

More generally, suppose Alice and Bob share $n$ qudit pairs of the form $\sum_{i=1}^N \alpha_i\ket{i}\ket{i}$.
There is still a linear dependence whenever $n > N$, no matter what the coefficients $\alpha_i$ are.
Specifically, the superposition
\begin{equation}
	\ket{\psi}=\sum_{\pi\in S_n}\sigma(\pi)\ket{\pi},
\end{equation}
where $\sigma(\pi)$ is the sign of the permutation $\pi$, is null when $n > N$.
As we will explain in more detail later, $\braket{\psi}$ is proportional to the sum
\begin{equation}
	\sum_{i_1<i_2<\cdots<i_n}\alpha_{i_1}^2\alpha_{i_2}^2\cdots\alpha_{i_n}^2
\end{equation}
which contains no terms when $n > N$.

In summary, for sufficiently large $n$ there is a nontrivial null state made up of superpositions of different permutation states.
By symmetry under $S_n$, the existence of such a null state implies that the unpermuted state $\ket{(1)}$ is equivalent to some superposition of permuted states.

\section{Shared thermofield double states}\label{TFD}
In the context of the AdS/CFT correspondence, the thermofield double state $\ket{\text{TFD}} = \sum_ie^{-\beta/2 E_i}\ket{E_i}\ket{E_i}$ is dual to a two-sided Schwarzschild black hole with inverse temperature $\beta$.
The density matrix for one side of the thermofield double (TFD) state describes the exterior on one side of the black hole, whereas the full state including the entanglement information between the two sides describes the full geometry including the interior of the black hole \cite{VanRaamsdonk:2016exw}.

Suppose Alice and Bob share $n$ such thermofield double states, but neither knows which of their $n$ systems is entangled with which of the other's.
This is analogous to the situation described in the previous section but now with an additional interpretation in terms of bulk geometries:
Alice and Bob each have access to $n$ black hole exteriors.
Each of Alice's exteriors is connected to one of Bob's via a black hole interior, but they do not know which is connected to which.
Alice and Bob are unable to perfectly determine which of the $n!$ geometries they have.
What's more, as we will argue, a linear dependency emerges as $n$ grows larger, so that any one situation can be approximately written as a linear combination of the others.

To be more precise, consider a boundary theory on a spatial manifold $\Sigma$.
Let $\ket{\text{TFD}(\beta)}\in H_L\otimes H_R$ be the thermofield double state in this theory with inverse temperature $\beta$.
Now consider $n$ copies of this thermofield double, i.e.\ the state $\ket{\text{TFD}(\beta)}\ket{\text{TFD}(\beta)}\cdots\ket{\text{TFD}(\beta)} = \ket{\text{TFD}(\beta)}^{\otimes n}$ in the Hilbert space $H_L^{\otimes n}\otimes H_R^{\otimes n}$.
We can act on this state $\ket{\text{TFD}(\beta)}^{\otimes n}$ by permuting the right boundaries, in other words by permuting the $n$ copies of the Hilbert space $H_R$.
There are $n!$ possible such boundary permutations, described by the symmetric group $S_n$.
Denote the state resulting from a permutation $\pi\in S_n$ as $\ket{\pi}\equiv\pi\circ\ket{\text{TFD}(\beta)}^{\otimes n}$.
\begin{figure}[t]
	\centering
	\includegraphics[width=0.5\linewidth]{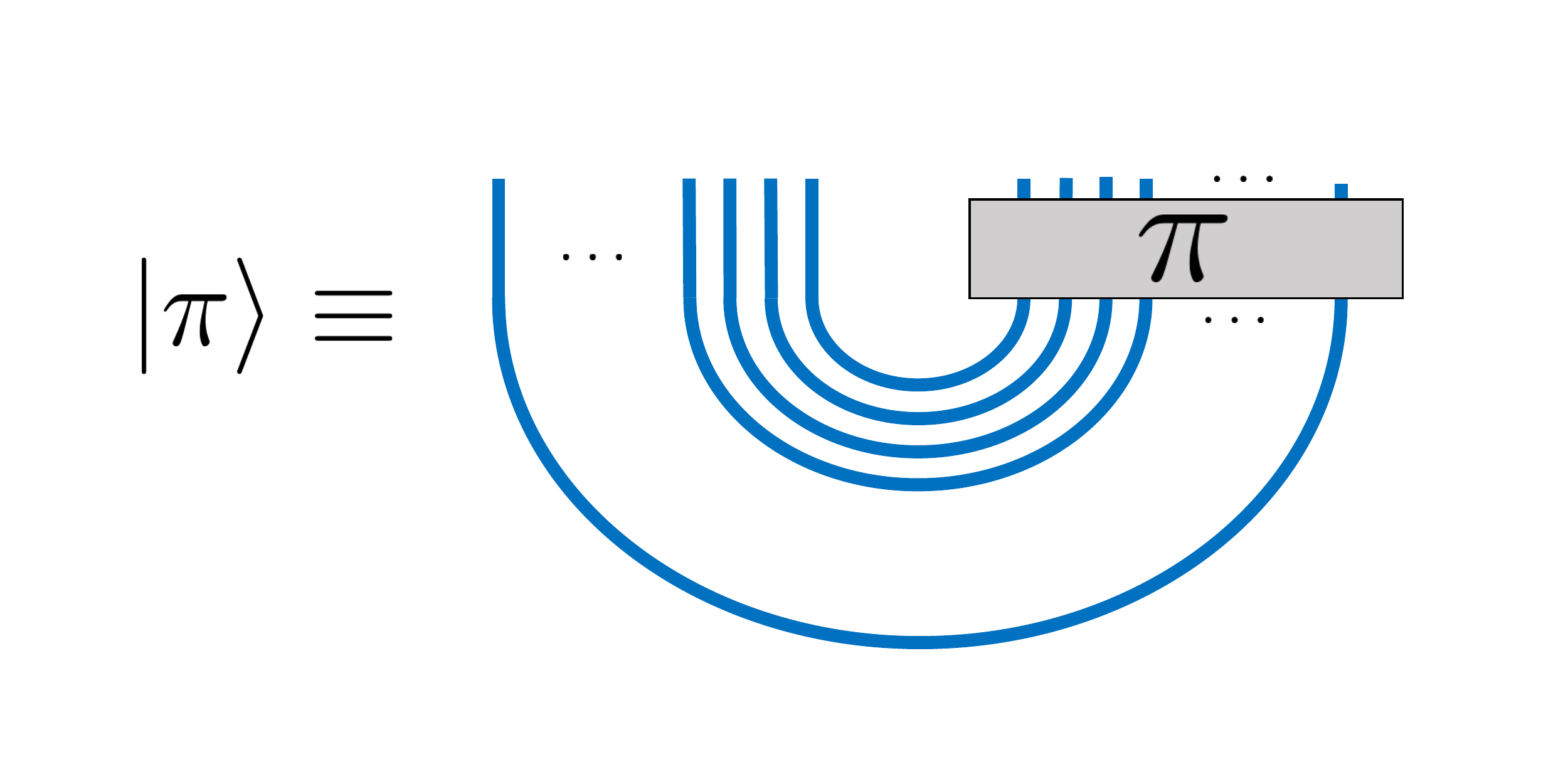}
	\caption{
		The state $\ket{pi}$
	}\label{fig:TFDstates_pi}
\end{figure}

First, note that the inner product between two states, $\ket{\pi}$ and $\ket{\t{\pi}}$, is $\braket{\t{\pi}}{\pi}=\braket{\text{id}}{\t{\pi}^{-1}\pi}=\expval{\t{\pi}^{-1}\pi}$, where for notational simplicity we denote inner products $\braket{\text{id}}{\pi}$ by $\expval{\pi}$.
\begin{figure}[t]
	\centering
	\begin{equation}
		\braket{\mathrm{id}}{(1\ 2)}
		\quad\quad\quad\quad
		=
		\vcenter{\hbox{\includegraphics[width=0.4\linewidth]{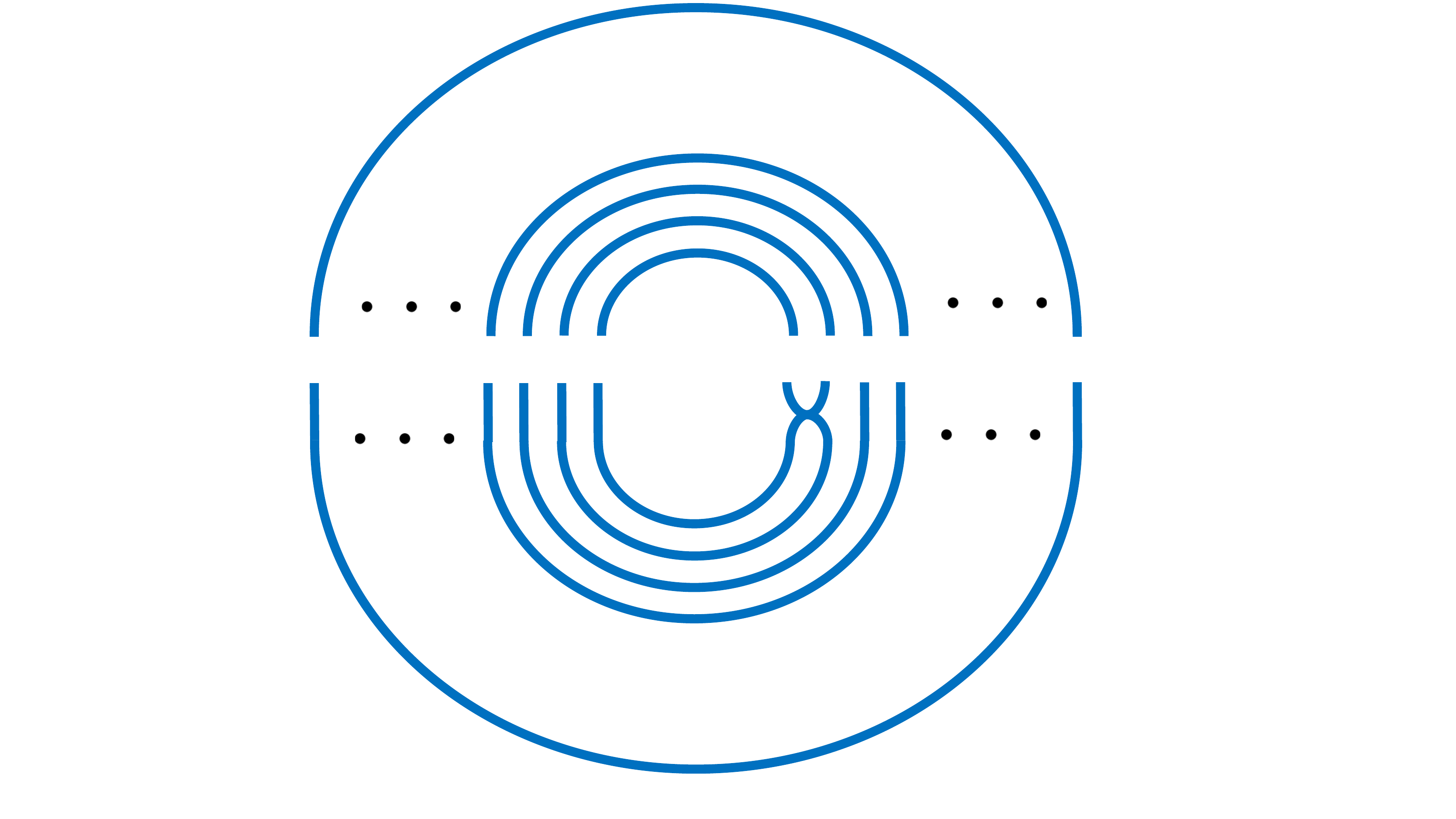}}}
	\end{equation}
	\caption{
		The inner product $\braket{\mathrm{id}}{(1\ 2)}=\expval{(1\ 2)}$.
		After normalization, this is $Z(2\beta)/Z(\beta)^2$.
	}
	\label{fig:TFDstates_overlap}
\end{figure}
Let $Z(\beta)$ be the partition function on $\Sigma\times S^1$ where the circle has length $\beta$.
Then we can evaluate $\expval{\pi}$ as 
\begin{equation}\label{expvaldefinition}
	\expval{\pi}=\prod_{\text{cycles $i$ of $\pi$}}Z(n_i\beta)
\end{equation}
where $n_i$ are the sizes of the cycles.
E.g.\ $\sum_i n_i=n$.
Note for later that $\expval{\pi}=\expval{\pi^{-1}}$.

The overlaps between these states will tend to be small.
The largest overlap, $\braket{\text{id}}{(1\ 2)}$, will be smaller than the norms $\braket{\text{id}}=Z(\beta)^n$ by a factor $Z(2\beta)/Z(\beta)^2$, which will be small on the order of the typical relative spacing $(e^{-\beta E_i} - e^{-\beta E_{i+1}})/e^{-\beta E_i} \approx \beta (E_{i+1}-E_i)$.
The smallness of the overlaps implies that the question of which sides of the TFD states are connected to which is \emph{approximately} a quantum observable, and that this could break down for large superpositions of such permutation states.\footnotemark
\footnotetext{In the special case where there is a Hagedorn temperature $1/\beta_*$, exact orthogonality is achieved in the limit $\beta\rightarrow \beta_*$, as $Z(2\beta)/Z(\beta)^2$ will become zero.}
In section \ref{example} we saw that for finite dimensional systems a counting argument implies linear dependence between the $n!$ permutation states for sufficiently large $n$.
In the more general case of infinite dimensional Hilbert spaces, the counting argument no longer holds.
To the extent that we can approximate a TFD state (and its dual black hole state) by truncating energies above a threshold, we can again obtain a null state by having sufficiently large $n$.
This suggests the possibility of obtaining a null state in the limit of large $n$, or a state whose difference from the null state (in a way we will make precise) goes to 0 as $n$ goes to infinity.
One goal of this work is to get a handle on such ``approximate null states."

A general superposition of $n$ permuted thermofield doubles is $\ket{A}=\sum_\pi\alpha_\pi\ket{\pi}$ and has norm
\begin{equation}
	\braket{A}{A}=\sum_\pi\sum_{\t{\pi}}\bar{\alpha}_{\t{\pi}}\alpha_\pi\braket{\t{\pi}}{\pi}=\sum_\pi\sum_{\t{\pi}}\bar{\alpha}_{\t{\pi}}\alpha_\pi M_{\t{\pi}\pi}
\end{equation}
where $M$ is the matrix with entries $M_{\t{\pi}\pi}=\braket{\t{\pi}}{\pi}=\expval{\t{\pi}^{-1}\pi}$.
Null states made up of the states $\ket{\pi}$ will correspond to eigenvectors of $M$ with eigenvalue $0$.

It isn't hard to find the eigenvalues of $M$, given its particular structure in terms of the group $S_n$. Let $U^{(q)}(\pi)$ be be the irreps of $S_n$, labeled by $q$.
The eigenstates of $M$ are then the states of the form
\begin{equation}
	\ket{q;ij}\equiv\sum_{\pi}U^{(q)}_{ij}(\pi)\ket{\pi},
\end{equation}
where $U^{(q)}_{ij}(\pi)$ is the $ij$ entry of the matrix $U^{(q)}(\pi)$.
This is simple to check using Schur orthogonality.
First note that $\expval{\pi}$ is a class function, so it can be uniquely expanded as a sum of characters $\chi^q$ of $S_n$:
\begin{equation}
	\expval{\pi}=\sum_q c_q\chi^q(\pi).
\end{equation}
Then
\begin{equation}
	\begin{aligned}
		\braket{p;kl}{q;ij}
		&=
		\sum_\pi\sum_{\t{\pi}}
		\overline{U^{(p)}_{kl}(\t{\pi})}
		U^{(q)}_{ij}(\pi)
		\expval{\t{\pi}^{-1}\pi}\\
		&=
		\sum_rc_r\sum_\pi\sum_{\t{\pi}}
		\overline{U^{(p)}_{kl}(\t{\pi})}
		U^{(q)}_{ij}(\pi)
		\sum_mU^{(r)}_{mm}(\pi^{-1}\t{\pi})\\
		&=
		\sum_rc_r\sum_\pi\sum_{\t{\pi}}
		\overline{U^{(p)}_{kl}(\t{\pi})}
		U^{(q)}_{ij}(\pi)
		\sum_m\sum_h
		U^{(r)}_{mh}(\pi^{-1})
		U^{(r)}_{hm}(\t{\pi})\\
		&=
		\sum_m\sum_h\sum_rc_r
		\sum_{\t{\pi}}
		\overline{U^{(p)}_{kl}(\t{\pi})}
		U^{(r)}_{hm}(\t{\pi})
		\sum_{\pi}
		U^{(q)}_{ij}(\pi)
		\overline{U^{(r)}_{hm}(\pi)}\\
		&=
		\sum_m\sum_h\sum_rc_r
		\frac{n!}{d_p}\delta_{pr}\delta_{kh}\delta_{lm}
		\frac{n!}{d_q}\delta_{qr}\delta_{ih}\delta_{jm}
		\\
		&=
		\delta_{pq}\delta_{ki}\delta_{lj}\frac{n!^2}{d_q^2}c_q,
	\end{aligned}
\end{equation}
where $d_q$ is the dimension of the irrep $q$.
We used Schur orthogonality between the fourth and fifth lines, and the fact $\expval{\pi}=\expval{\pi^{-1}}$ between the first and second lines.
So the states $\ket{q;ij}$ are orthogonal, with norm-squared $\frac{n!^2}{d_q^2}c_q$.
They give an orthogonal basis for the space spanned by the $\ket{\pi}$ states.

This means that any null states will be precisely those $\ket{q;ij}$ for which $c_q$ is 0.
By their definition, the coefficients $c_q$ are
\begin{equation}
	c_q=\frac{1}{n!}\sum_{\pi}\chi^q(\pi)\expval{\pi}.
\end{equation}
In general, for $q$ other than the trivial representation, the values $\chi^q(\pi)$ can be either positive negative.
So we can't rule out null states a priori, even if we do not expect them exactly for finite $n$ in an infinite dimensional Hilbert space.
We will tend to assume that the $c_q$ are nonzero.

It is worth pointing out, that the coefficients $c_q$ are the Schur polynomials \cite{Wiki:Schur} (for finite dimensional systems, Schur functions more generally) in the variables $e^{-\beta E_1}$, $e^{-\beta E_2}$, \ldots .
Specifically, the Schur function $s_\lambda(e^{-\beta E_1}, e^{-\beta E_2}, \ldots)$ for a partition $\lambda$ of $n$ is the coefficient $c_q$ for the $S_n$ irrep corresponding to $\lambda$.
In Section \ref{example} we have already used the fact that $c_\sigma=c_{(1,1,\ldots,1)}$, the coefficient corresponding to the sign irrep $\sigma$ of $S_n$, is the Schur polynomial $s_{(1,1,\ldots,1)}$ corresponding to the partition $1+\ldots +1=n$, together with the fact that the Schur polynomial $s_{(1,1,\ldots, 1)}(x_1,x_2,\ldots)$ equals the $n$-th elementary polynomial
\begin{equation}
	e_n(x_1,x_2,\ldots)\equiv \sum_{i_1<\cdots<i_n} x_{i_1}\cdots x_{i_n}.
\end{equation}
For simplicity and clarity we will sometimes notate partitions with ``exponential notation", so that e.g.\ $c_{(1,1,\ldots,1)}$ is written $c_{(1^n)}$.
We will find the fact
\begin{equation}
	c_{(1^n)} = \sum_{i_1<\cdots<i_n} e^{-\beta(E_{i_1}+\cdots+E_{i_n})}
\end{equation}
to be of importance later.
One additional fact about Schur polynomials we will use is that
\begin{equation}\label{Schurrelation}
	\left(\sum_i x_i\right)s_\lambda(x_1,x_2,\ldots) = \sum_\mu s_\mu(x_1,x_2,\ldots),
\end{equation}
where the sum on the RHS is over partitions $\mu$ obtainable by adding a single box to the young diagram of $\lambda$.

\subsection{Nearly null states}\label{nearlynullstates}
On it's face it does not make much sense to refer to a state as ``almost" null, as any state whose norm is not zero can be normalized to have norm 1.
To define a sense of ``nearly null" states, or even a sequence of states whose limit is the null state we must have some additional criterion that determines a preferred normalization.
A nearly null state is then one whose preferred normalization, in this sense, is small.

One outcome of having a null state, say $\ket{a}-\ket{b}=0$, is that we can substitute $\ket{b}$ for $\ket{a}$ in expressions and get the same result.
We can take an ``approximate null state" $\ket{a}-\ket{b}\approx 0$ to include the criterion that replacing $\ket{a}$ with $\ket{b}$ gives us \emph{approximately} the same result.
For example, we could require that the fidelity between $\ket{a}$ and $\ket{b}$ be close to one.
This is equivalent to the norm of $\ket{a}-\lambda\ket{b}$ being small relative to the norm of $\ket{a}$ for some choice of $\lambda$.

We are interested in states of the form $\ket{\text{geometry}} + \sum_i\alpha_i\ket{\text{other geometry $i$}}$ and the possibility $\ket{\text{geometry}} \approx - \sum_i\alpha_i\ket{\text{other geometry $i$}}$.
So the notion of ``nearly null" we will be interested in is states with norm that is small compared to the norm of a distinguished term $\ket{\text{geometry}}$ or states $\ket{\phi}$ such that the fidelity between $\ket{\text{geometry}}$ and $\ket{\phi}-\ket{\text{geometry}}$ is close to 1.
This definition of course only makes sense relative to some distinguished vector or vectors.
In our context we have the distinguished vector $\ket{\mathrm{id}}=\ket{\text{TFD}}^{\otimes n}$ and the states obtained by permutations acting on this.

First we consider the possibility of writing the state of $n$ unpermuted two-sided black holes, $\ket{\text{id}}$, as a superposition $\ket{\psi}=\sum_{\pi\neq \text{id}}\alpha_\pi\ket{\pi}$ of permuted states.
We wish to find the maximum fidelity between $\ket{\text{id}}$ and $\ket{\psi}$:
\begin{equation}\label{quant1}
	\begin{aligned}
		F_{\text{max}}(n)=\max_{\ket{\psi}}\frac{\left|\braket{\mathrm{id}}{\psi}\right|^2}{\braket{\mathrm{id}}{\mathrm{id}}\braket{\psi}{\psi}}
	\end{aligned}
\end{equation}
where the maximization is over states $\ket{\psi}$ that are linear combinations only of $\ket{\pi}$ where $\pi\neq\text{id}$.
We can evaluate \eqref{quant1} by expanding in the $\ket{q;ij}$ basis.
First we expand $\ket{\psi}$ as
\begin{equation}
	\ket{\psi}=\sum_{\pi\neq \mathrm{id}}\alpha_\pi\ket{\pi}=\sum_q\sum_{i,j=1}^{d_q}\gamma_{qij}\ket{q;ij}
\end{equation}
where $\gamma_{qij}$ are coefficients satisfying
\begin{equation}
	\alpha_\pi=\sum_q\sum_{i,j=1}^{d_q}\gamma_{qij}U^{(q)}_{ij}(\pi).
\end{equation}
In terms of the $\gamma_{qij}$ coefficients, the condition that $\alpha_{\mathrm{id}}$ be zero is given by
\begin{equation}
	\sum_q\sum_{i=1}^{d_q}\gamma_{qii}=0.
\end{equation}
Now note that
\begin{equation}
	\begin{aligned}
		\braket{\text{id}}{q;ij}
		&=
		\sum_{\pi}U^{(q)}_{ij}(\pi)\expval{\pi}\\
		&=
		\sum_\pi U^{(q)}_{ij}(\pi)\left(\sum_pc_p\chi_p(\pi)\right)\\
		&=
		\sum_k\sum_pc_p\sum_\pi U^{(q)}_{ij}(\pi)U^{(p)}_{kk}(\pi)\\
		&=
		\sum_k\sum_pc_p\delta_{i,k}\delta_{j,k}\delta_{\bar{q},p}\frac{n!}{d_p}\\
		&=\delta_{ij}\frac{n!}{d_q}c_q
	\end{aligned}
\end{equation}
where we have used Schur orthogonality as well as the fact that the $c_q$ are real.
This fact now gives us
\begin{equation}
	\begin{aligned}
		\braket{\mathrm{id}}{\psi}
		&=
		\sum_q\sum_{i=1}^{d_q}\gamma_{qii}\frac{n!}{d_q}c_q
	\end{aligned}
\end{equation}
as well as
\begin{equation}
	\begin{aligned}
		\braket{\psi}{\psi}
		&=
		\sum_q\sum_{i,j=1}^{d_q}\overline{\gamma_{qij}}\gamma_{qij}\frac{n!^2}{d_q^2}c_q.
	\end{aligned}
\end{equation}
Plugging these expressions into \eqref{quant1}, we get
\begin{equation}\label{quant4}
	\begin{aligned}
		F_{\text{max}}=\max_{\gamma}\frac{\left|\sum_q\frac{n!}{d_q}c_q\tr\gamma_{q}\right|^2}{Z^n\sum_q\frac{n!^2}{d_q^2}c_q\tr(\gamma_{q}^\dag\gamma_{q})}.
	\end{aligned}
\end{equation}
where the maximization is subject to the constraint that $\sum_q\sum_{i=1}^{d_q}\gamma_{qii}=0$, and $\gamma_q$ denotes the $d_q$-by-$d_q$ matrix whose entries are $\gamma_{qij}$.
Splitting the matrix $\gamma_q$ up into its trace and traceless parts, $\gamma_q=\frac{\tr\gamma_q}{d_q}\1+X$ we see that $\tr(\gamma_q^\dag\gamma_q)=\frac{1}{d_q}\tr(\gamma_q)^2+\tr(X^\dag X)$. The traceless degrees of freedom of $\gamma_q$ do not participate in either the constraint, or the numerator of \eqref{quant4}. They do add a nonnegative contribution to the denominator of \eqref{quant4}, however, which can only decrease $F$. So, because they don't participate in the constraint, they will be $0$ when $F$ is maximized. Define
\begin{equation}
	\lambda_q=\sqrt{Z^n\frac{n!^2 c_q}{d_q^3}}\tr\gamma_q.
\end{equation}
With the traceless part of $\gamma_q$ set to zero the max fidelity is now
\begin{align}
	F_{\text{max}}=\max_{\lambda}\frac{\left|\sum_q\sqrt{\frac{d_qc_q}{Z^n}}\lambda_q\right|^2}{\sum_q\left|\lambda_q\right|^2}
\end{align}
where the maximization is over the $\lambda_q$ and the constraint can be written
\begin{equation}
	\sum_q\sqrt{\frac{d_q^3}{Z^n n!^2c_q}}\lambda_q=0.
\end{equation}
Written this way, $F_{\text{max}}$ is simply the maximum (normalized) overlap between the vectors $\vec{u}=\left(\ldots,\sqrt{\frac{d_qc_q}{Z^n}},\ldots\right)$ and $\vec{\lambda}=\left(\ldots,\lambda_q,\ldots\right)$ with the constraint that $\vec{\lambda}$ is orthogonal to the vector $\vec{v}=\left(\ldots,\sqrt{\frac{d_q^3}{Z^n n!^2c_q}},\ldots\right)$. You can then see that the maximum is obtained by choosing $\vec{\lambda}$ in the plane defined by $\vec{u}$ and $\vec{v}$ and orthogonal to $\vec{\lambda}$.
Where $\theta$ is the angle between $\vec{u}$ and $\vec{v}$, the angle between the optimal $\vec{\lambda}$ and $\vec{v}$ is $\frac{\pi}{2}-\theta$, making the maximum fidelity
$F_{\text{max}}=\left(\cos(\frac{\pi}{2}-\theta)\right)^2=1-(\cos\theta)^2$.
To easily get $\theta$:
\begin{equation}
	\cos\theta=\frac{\vec{u}\cdot\vec{v}}{\left\|u\right\|\left\|v\right\|}=\frac{\sum_q\frac{d_q^2}{Z^n n!}}{\sqrt{\sum_q\frac{d_qc_q}{Z^n}}\sqrt{\sum_q\frac{d_q^3}{Z^n n!^2 c_q}}}.
\end{equation}
After simplifications from the facts $\sum_q d_q^2=n!$ and $\sum_q d_qc_q=Z(\beta)^n$,\footnotemark this gives the maximum fidelity
\footnotetext{The second fact here can be seen by expanding the definitions of the $c_q$ then noting that the LHS is $\frac{1}{n!}\sum_\pi\chi_{\text{reg}}(\pi)\expval{\pi}$, where $\chi_{\text{reg}}(\pi)=n!\delta_{\mathrm{id},\pi}$ is the regular representation.}
\begin{equation}\label{nearlynull}
	F_{\text{max}}=1-\frac{n!^2}{Z(\beta)^n\sum_q\frac{d_q^3}{c_q}}.
\end{equation}
We remind the reader that the sum is over irreps $q$ of $S_n$ so that the sum, the dimensions $d_q$ and the coefficients $c_q$ all have a dependence on $n$.

This result gives us a criterion for having an approximate null state: is $n!^2/\left(Z^n\sum_qd_q^3/c_q\right)$ small?
Of course determining the coefficients $c_q$ or even getting a bound on them may be difficult in general.
If any of $c_q(n)/Z^n$ go to zero as $n$ gets large, the fidelity goes to 1.
This is also consistent with the fact already seen that there is an exact null state whenever one of the $c_q$ is zero.
Further note that in the above derivation of $F_{\text{max}}$ we have not used any assumptions about the partition function, in particular we have not assumed either finite dimensions or a discrete spectrum.

We can go further and obtain an actual state that instantiates the above maximum fidelity.
This will be the projection of $\vec{u}$ to the subspace orthogonal to $\vec{v}$.
So
\begin{equation}
	\begin{aligned}
		\vec{\lambda}_{\text{max}}
		&\sim
		\vec{u}-\frac{\vec{v}\cdot\vec{u}}{\left\|v\right\|^2}\vec{v}.
	\end{aligned}
\end{equation}
Unpacking our definition of $\lambda_q$ in terms of $\tr(\gamma_q)$ we get
\begin{equation}
	\tr(\gamma_q)
	\sim
	\frac{d_q^2}{Z^nn!}-\frac{d_q^3/c_q}{Z^n\sum_p d_p^3/c_p}
\end{equation}
after some simplification.
This gives the coefficients
\begin{equation}
	\gamma_{qij}
	\sim
	\delta_{ij}\left(\frac{d_q}{Z^nn!}-\frac{d_q^2/c_q}{Z^n\sum_p d_p^3/c_p}\right).
\end{equation}
This result, in terms of the original coefficients $\alpha_\pi$, is
\begin{equation}\label{nearlynullalpha}
	\begin{aligned}
		\alpha_\pi
		&\sim
		\sum_q\sum_{i,j=1}^{d_q}
		\delta_{ij}\left(\frac{d_q}{Z^nn!}-\frac{d_q^2/c_q}{Z^n\sum_p d_p^3/c_p}\right)
		U^{(q)}_{ij}(\pi)\\
		&=
		\frac{\chi_\text{reg}(\pi)}{Z^nn!}-\frac{1}{Z^n}\frac{\sum_q d_q^2\chi^q(\pi)/c_q}{\sum_p d_p^3/c_p}\\
		&\sim
		\delta_{\pi,\mathrm{id}}-\frac{\sum_q d_q^2\,\chi^q(\pi)/c_q}{\sum_q d_q^2\,\chi^q(\mathrm{id})/c_q}.
	\end{aligned}
\end{equation}
Written this way, we can see that $\alpha_\mathrm{id}$ is indeed $0$, consistent with the constraint on the optimization.

We have determined the optimal $\ket{\psi}$ up to an overall constant.
To fix the constant, remember that we are interested in interpreting the state $\ket{\mathrm{id}}-\ket{\psi}$ as a ``nearly null" state.
We choose the phase of $\ket{\psi}$ so that $\braket{\mathrm{id}}{\psi}$ is real and positive, and we choose the norm to be close to the norm of $\ket{\mathrm{id}}$.
Taking $\alpha_\pi$ to be equal to the last line of \eqref{nearlynullalpha} gives
\begin{equation}
	\braket{\mathrm{id}}{\psi}=Z^nF_{\text{max}},
\end{equation}
which is positive.
And likewise
\begin{equation}
	\braket{\psi}{\psi}=Z^nF_{\text{max}}
\end{equation}
which for $F_{\text{max}}$ close to 1 as desired will be approximately $Z^n=\braket{\mathrm{id}}$.
Thus, granted $F_{\text{max}}$ is close to 1, we get a suitable ``nearly null" state in
\begin{equation}
	\ket{\mathrm{id}}-\ket{\psi}=\ket{\mathrm{id}}+\sum_\pi\left(\frac{\sum_q\frac{d_q^2}{c_q}\chi^q(\pi)}{\sum_q\frac{d_q^2}{c_q}\chi^q(\mathrm{id})}-\delta_{\pi,\mathrm{id}}\right)\ket{\pi}
\end{equation}
or, written more simply,
\begin{equation}
	\ket{\text{nearly null}}\sim\sum_{\pi\in S_n}\left(\sum_q\frac{d_q^2}{c_q}\chi^q(\pi)\right)\ket{\pi}.
\end{equation}

\subsection{Meeting in a black hole interior}
In the previous Section \ref{nearlynullstates} we found a candidate state $\ket{\mathrm{id}}-\sum_{\pi\neq\mathrm{id}}\alpha_\pi\ket{\pi}$ for the property $\ket{\mathrm{id}}\approx \sum_{\pi\neq\mathrm{id}}\alpha_\pi\ket{\pi}$.
Suppose this were an equality.
Then by combining on one side the terms $\ket{\pi}$ where $\pi(1) = 1$, we could write this in the form
\begin{equation}\label{nearlynullstaterearranged}
	\ket{\text{TFD}}
	\otimes
	\sum_{
		\t{\pi}\in S_{n-1}
	}
	\alpha_{\t{\pi}}\ket{\t{\pi}}
	=
	-
	\sum_{
		\substack{
			\pi\in S_n\\
			\pi(1)\neq 1
		}
	}
	\alpha_\pi\ket{\pi}.
\end{equation}
In a holographic context, the RHS would be a superposition of geometries, none of which connect the first black hole exterior on each side, but whose partial trace down to the first black hole is a connected black hole geometry.\footnotemark
\footnotetext{There is also the possibility that $\sum_{\t{\pi}\in S_{n-1}}\alpha_{\t{\pi}}\ket{\t{\pi}}$ is itself null.
	If this is the case, we apply the same consideration to it, extracting terms that have $\ket{\text{TFD}}$ on the \emph{second} black hole, and so on.
	This process will bottom out eventually, as we have fixed the coefficient of $\ket{\mathrm{id}}$ to be 1.}
Two observers jumping into the first black hole could in principle meet in the middle, this is despite the fact in no term in the RHS of \eqref{nearlynullstaterearranged} has a geometry that connects the observers.

We extend our search to superpositions of states $\ket{\pi}$ with $\pi(1)\neq 1$ whose partial trace down to the first copy of $H_L\otimes H_R$ is approximately the thermofield double state.
This previous paragraph shows that this is a weaker condition than that considered in Section \ref{nearlynullstates}, and it will turn out to be easier to get good bounds on the fidelity for this case.
So consider a superposition $\ket{\psi} = \sum_{\pi, \pi(1)\neq 1}\alpha_\pi\ket{\pi}$ and its partial trace $\rho_1 = \tr_{2\ldots n}\big(\ketbra{\psi}\big)$.
We want to maximize the fidelity
\begin{equation}\label{TFDfidelity}
	F=
	\frac{\bra{\text{TFD}}\rho_1\ket{\text{TFD}}}{\braket{\text{TFD}}\tr(\rho_1)}
	=
	\frac{\bra{\psi}\big(\ketbra{\text{TFD}}\otimes\1_{2\ldots n}\big)\ket{\psi}}{\braket{\text{TFD}}\braket{\psi}}
\end{equation}
Like before, we will expand this in the irrep basis $\ket{q;ij}$, though this time we will consider both the basis $\ket{q;ij}_{1\cdots n}=\sum_{\pi\in S_n}U^{(q)}_{ij}(\pi)\ket{\pi}_{1\cdots n}$ for $n$ copies of $H_L\otimes H_R$ and the basis $\ket{p;k\ell}_{2\cdots n}=\sum_{\pi\in S_{n-1}}U^{(p)}_{k\ell}(\pi)\ket{\pi}_{2\cdots n}$ for only the the last $n-1$ copies of $H_L\otimes H_R$.
We will allow ourselves to drop the subscripts when there is no ambiguity and will tend to use $q$ for irreps of $S_n$ and $p$ for irreps of $S_{n-1}$ in what follows.

Let $P=\sum_{p,k,\ell}\ketbra{p;k\ell}/\braket{p;k\ell}$ be the projector onto the subspace spanned by the states $\ket{\pi}_{2\cdots n}$ with $\pi\in S_{n-1}$.
Replacing the identity with $P$ in the expression for fidelity gives a lower bound
\begin{equation}\label{projectionbound}
	F
	\geq
	\frac{
		\bra{\psi}\big(\ketbra{\text{TFD}}\otimes P\big)\ket{\psi}
	}{
		\braket{\text{TFD}}\braket{\psi}
	}.
\end{equation}
The norms $\braket{p;k\ell}$ are $n!^2c_p/d_p^2$.
Expanding $\ket{\psi}=\sum_{q,i,j}\gamma_{qij}\ket{q;ij}$ and $P$, we get an expression involving inner products $\big(\bra{\text{TFD}}\bra{p;k\ell}\big)\ket{q;ij}$.
These can be worked out.
An irrep $q$ of $S_n$ will split when restricted to $S_{n-1}$ into those irreps whose young diagrams are obtained by removing a single corner block from the young diagram of $q$.
Assume we chose our bases such that the matrices $U^{(q)}_{ij}(\pi)$ for $S_n$ irrep $q$ are block diagonal when restricted to $\pi\in S_{n_1}$ and so that the indices $k$ and $\ell$ match an appropriate subset of the indices $i$ and $j$, when $p$ is a block in the restriction of $q$.
We have
\begin{equation}
	\begin{aligned}
		\big(\bra{\text{TFD}}\otimes\bra{p;k\ell}\big)\ket{q;ij}
		&=
		\sum_{\t{\pi}\in S_{n-1}}\sum_{\pi\in S_n}
		\b{U^{(p)}_{k\ell}(\t{\pi})}
		U^{(q)}_{ij}(\pi)
		\braket{\t{\pi}}{\pi}\\
		&=
		\sum_m\sum_{\t{\pi}\in S_{n-1}}\sum_{\pi\in S_n}
		\b{U^{(p)}_{k\ell}(\t{\pi})}
		U^{(q)}_{im}(\t{\pi})U^{(q)}_{mj}(\pi)
		\expval{\pi}.
	\end{aligned}
\end{equation}
Expanding $U^{(q)}_{im}(\t{\pi})$ into blocks and using Schur orthogonality we obtain
\begin{equation}
	\big(\bra{\text{TFD}}\otimes\bra{p;k\ell}\big)\ket{q;ij}
	=
	\frac{(n-1)!}{d_p}\delta_{p\subset q}\delta_{ki}\sum_{\pi\in S_n}U^{(q)}_{\ell j}(\pi)\expval{\pi},
\end{equation}
where by $\delta_{p\subset q}$ we mean 1 when $p$ is in the restriction of $q$ and 0 if it is not, and $\delta_{ki}$ is meant to be understood as enforcing that the $k$-th index of the copy of $p$ in $q$ matches the $i$-th index of $q$.
The remaining sum here evaluates to $\delta_{\ell j}n!c_q/d_q$, which can be seen by expanding $\expval{\pi}$ into characters of $S_n$ then using Schur orthogonality.
So in all we have
\begin{equation}
	\big(\bra{\text{TFD}}\otimes\bra{p;k\ell}\big)\ket{q;ij}
	=
	\delta_{p\subset q}\delta_{ki}\delta_{\ell j}\frac{(n-1)!}{d_p}\frac{n!}{d_q}c_q.
\end{equation}

After expanding $\ket{\psi}$ and $P$ into the $\ket{q;ij}$ and $\ket{p;k\ell}$ bases respectively, our bound \eqref{projectionbound} becomes
\begin{equation}\label{expandedbound}
	F
	\geq
	\frac{
		\sum_{p,k,\ell}\sum_{q\supset p}\sum_{\t{q}\supset p}
		\b{\gamma_{\t{q}k\ell}}\gamma_{qk\ell}
		\frac{c_{\t{q}}}{d_{\t{q}}}
		\frac{c_q}{d_q}
		\frac{1}{c_p}
	}{
		Z(\beta)\sum_{p,k,\ell}\sum_{q\supset p}
		\left|\gamma_{qk\ell}\right|^2\frac{c_q}{d_q^2}
	}.
\end{equation}
The constraint on $\ket{\psi}$ that $\alpha_\pi$ be zero for all $\pi$ such that $\pi(1)=1$ becomes
\begin{equation}
	0 = \sum_{qij}\gamma_{qij}U^{(q)}_{ij}(\pi)\quad\text{for}\quad \pi\in S_{n-1}.
\end{equation}
Split into $p$ blocks this is
\begin{equation}
	0 = \sum_{pk\ell}\sum_{q\supset p}\gamma_{qk\ell}U^{(p)}_{k\ell}(\pi)\quad\text{for}\quad \pi\in S_{n-1}.
\end{equation}
Together, these conditions are equivalent to 
\begin{equation}
	0 = \sum_{q\supset p}\gamma_{qk\ell}\quad\text{for all $p$, $k$, and $\ell$},
\end{equation}
the constraint that the sum of all blocks in $\gamma_{qij}$ corresponding to an $S_{n-1}$ irrep must be the zero matrix.
Define $\lambda_{pk\ell q}=\frac{\sqrt{c_q}}{d_q}\gamma_{qk\ell}$ where, for a given $p$, $q$ ranges over irreps of $S_n$ that include the $S_{n-1}$ irrep $p$ in their restriction.
Then \eqref{expandedbound} becomes
\begin{equation}
	F
	\geq
	\frac{
		\sum_{p,k,\ell}
		\frac{1}{c_p}
		\sum_{\t{q}\supset p}
		\sqrt{c_{\t{q}}}
		\b{\lambda_{pk\ell \t{q}}}
		\sum_{q\supset p}
		\sqrt{c_q}
		\lambda_{pk\ell q}
	}{
		Z(\beta)\,\vec{\lambda}\cdot\vec{\lambda}
	}.
\end{equation}
Given fixed contribution $a_{pk\ell}^2=\sum_{q\supset p}\left|\lambda_{pk\ell q}\right|^2$ to the norm-squared from a given $p,k,\ell$ block, maximizing within that block is the problem of maximizing $\vec{u}\cdot \vec{v}$ given $\vec{w}\cdot \vec{v}=0$ where $\vec{v}$ has fixed norm $a_{pk\ell}$, where $v_q=\lambda_{pk\ell q}$, $u_q=\sqrt{c_q}$, and $w_q=d_q/\sqrt{c_q}$.
The maximizing $\vec{v}$ will be proportional the projection of the $\vec{u}$ to the subspace orthogonal to $\vec{w}$.
The result is
\begin{equation}
	F_{\text{max}}
	\geq
	\max_{a_{pk\ell}}
	\frac{
		\sum_{p,k,\ell}
		a_{pk\ell}^2
		\frac{1}{c_p}
		\left(
		\sum_{q\supset p}c_q
		-
		\frac{\left(\sum_{q\supset p}d_q\right)^2}{\sum_{q\supset p}d_q^2/c_q}
		\right)
	}{
		Z(\beta)\,\sum_{p,k,\ell}a_{pk\ell}^2
	}.
\end{equation}
Maximizing over the $a_{pk\ell}$ simply results in
\begin{equation}
	F_{\text{max}}
	\geq
	\max_p
	\frac{1}{Z(\beta)c_p}
	\left(
	\sum_{q\supset p}c_q
	-
	\frac{\left(\sum_{q\supset p}d_q\right)^2}{\sum_{q\supset p}d_q^2/c_q}
	\right).
\end{equation}
Using an aforementioned fact about Schur functions, $Z(\beta)c_p=\sum_{q\supset p}c_q$, this simplifies to
\begin{equation}\label{fidelityboundmaxp}
	F_{\text{max}}
	\geq
	1
	-
	\min_p
	\frac{\left(\sum_{q\supset p}d_q\right)^2}{\sum_{q\supset p}c_q\sum_{q\supset p}d_q^2/c_q}
	.
\end{equation}
Numerical experimentation suggests that the optimal irrep $p$ of $S_{n-1}$ is often the sign irrep.
This is not always the case however.
We will encounter a counterexample in Section \ref{semiclassical}.

In any case, choosing the sign irrep does give a lower bound on $F_{\text{max}}$.
There are two irreps $q$ of $S_n$ whose reduction to $S_{n-1}$ includes the sign irrep, namely the irreps corresponding to the partition $2+1+\cdots+1=n$ and to the sign irrep of $S_n$ (whose corresponding partition is $1+1+\cdots+1=n$).
The dimensions $d_q$ of these irreps are $n-1$ and $1$ respectively.
This leads to our final expression
\begin{equation}\label{fidelitybound}
	F_{\text{max}}
	\geq
	1 - 
	\frac{n^2}{1 + (n-1)^2 + \frac{c_{(2,1^{n-2})}}{c_{(1^n)}} + (n-1)^2\frac{c_{(1^n)}}{c_{(2,1^{n-2})}}}
\end{equation}
In examples, the coefficient $c_{(1,\ldots,1)}$ can be easier to calculate than other coefficients $c_q$.
It is thus sometimes useful to use $Z(\beta)c_{(1^{n-1})}=c_{(1^n)} + c_{(2,1^{n-2})}$ to write $c_{(2,1^{n-2})}$ in terms of the potentially easier to calculate $c_{(1^n)}$ and $c_{(1^{n-1})}$.
Similarly, $c_{(n)}$ and $c_{(n-1,1)}$ may be easier to calculate in examples than other coefficients.

\section{Harmonic oscillator}\label{harmonicoscillator}
To illustrate the application of the results of Section \ref{TFD} we consider the simple (though non-holographic) example of a harmonic oscillator.
In this case, the coefficients $c_{(1^n)}$ can be calculated using the recursion relation between elementary polynomials of different degree:
\begin{equation}
	e_n(x_1,x_2,\ldots) = e_n(x_2,x_3,\ldots) + x_1 e_{n-1}(x_2,x_3,\ldots).
\end{equation}
For a harmonic oscillator with spacing $\omega$ between energy levels, this gives
\begin{equation}
	\begin{aligned}
		c_{(1^n)}
		&=
		e_n(e^{-\beta\omega\frac{1}{2}},e^{-\beta\omega\frac{3}{2}},\ldots)\\
		&=
		e_n(e^{-\beta\omega\frac{3}{2}},e^{-\beta\omega\frac{5}{2}},\ldots)
		+
		e^{-\beta\omega\frac{1}{2}} e_{n-1}(e^{-\beta\omega\frac{3}{2}},e^{-\beta\omega\frac{5}{2}},\ldots)\\
		&=
		e^{-\beta\omega n} e_n(e^{-\beta\omega\frac{1}{2}},e^{-\beta\omega\frac{3}{2}},\ldots)
		+
		e^{-\beta\omega\frac{1}{2}}
		e^{-\beta\omega (n-1)}
		e_{n-1}(e^{-\beta\omega\frac{1}{2}},e^{-\beta\omega\frac{3}{2}},\ldots)\\
		&=
		e^{-\beta\omega n} c_{(1^n)}
		+
		e^{-\beta\omega (n-\frac{1}{2})}
		c_{(1^{n-1})}.
	\end{aligned}
\end{equation}
So the ratio $c_{(1^{n-1})}/c_{(1^n)}$ is
\begin{equation}
	\frac{c_{(1^{n-1})}}{c_{(1^n)}}
	=
	\frac{1-e^{-\beta\omega n}}{e^{-\beta\omega(n-\frac{1}{2})}}
	=
	e^{\beta\omega(n-\frac{1}{2})} - e^{-\beta\omega\frac{1}{2}}.
\end{equation}
With this we can calculate $c_{(2,1^{n-2})}=Z(\beta)\frac{c_{(1^{n-1})}}{c_{(1^n)}} - 1$ and apply \eqref{fidelitybound}.
The lower bound on the maximum fidelity is
\begin{equation}
	F_{\text{max}}
	\geq
	1
	-
	\frac{n^2}{
		\frac{e^{\beta\omega n}-1}{e^{\beta\omega}-1}
		+
		(n-1)^2
		\frac{e^{\beta\omega n}-1}{e^{\beta\omega n}-e^{\beta\omega}}
	}.
\end{equation}
This approaches 1 as $n$ gets large, as the first term in the denominator dominates for large $n$.
\begin{figure}[t]
	\centering
	\includegraphics[width=0.5\linewidth]{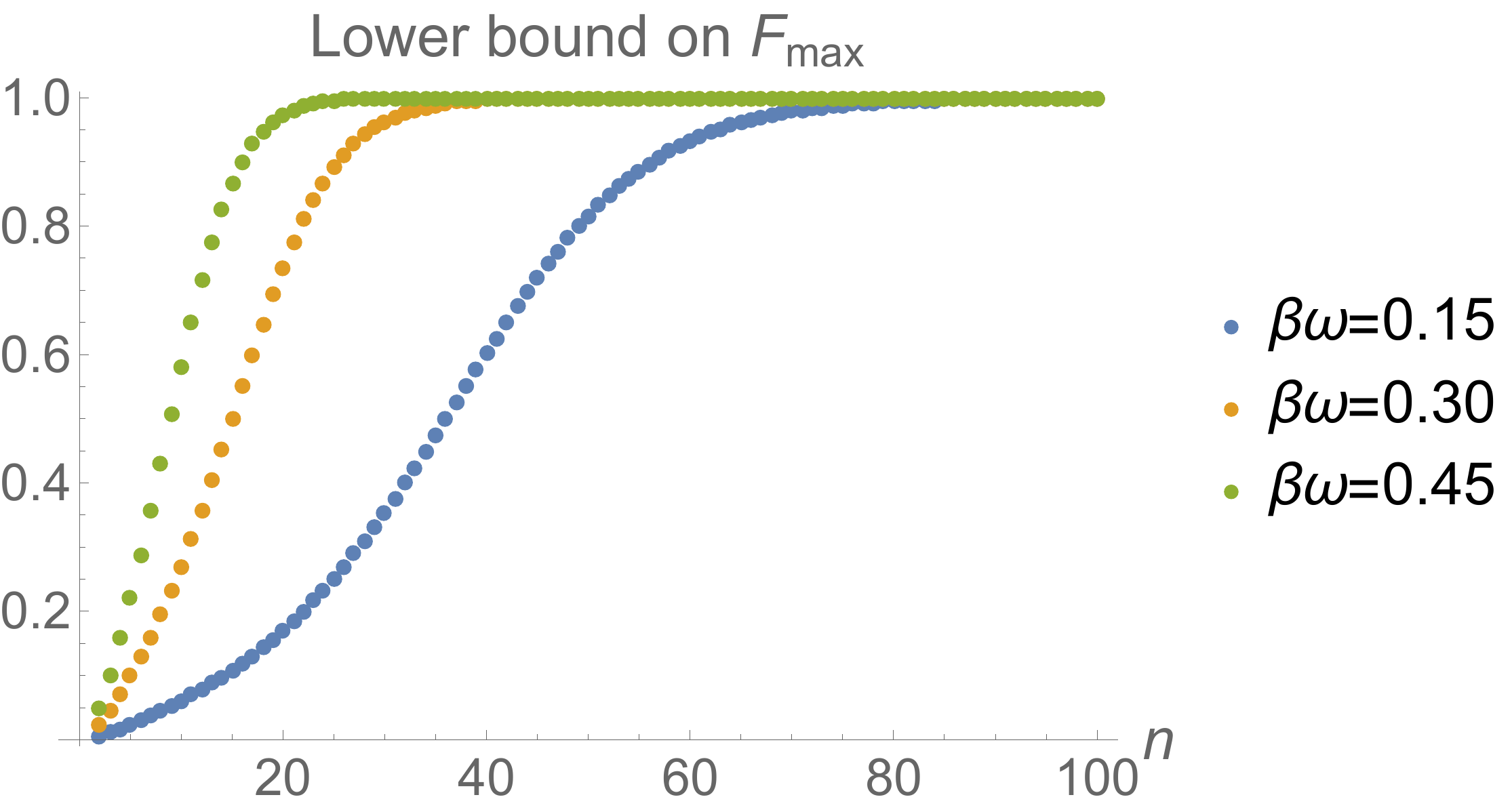}
	\caption{Lower bound on the maximum fidelity as a function of $n$, for harmonic oscillator TFD states with different values of $\beta\omega$}
	\label{fig:Harmonic_oscillator_F_versus_n}
\end{figure}
How quickly the fidelity approaches 1 as $n$ increases depends on $\beta\omega$ for the system, with smaller $\beta\omega$ giving a slower increase to 1.
This is general feature of other examples, that smaller spacing between energies or larger temperature both make it harder to achieve high fidelity.

\section{The Marolf and Maxfield toy model}
We now turn to a holographic example.
In \cite{Marolf:2020xie} the authors Marolf and Maxfield explore a simple toy model of a gravity path integral in 2 dimensions.
Interestingly, they show that this simple bulk path integral is holographically dual to an ensemble of 1 dimensional boundary theories, namely an ensemble of topological quantum mechanics theories where the dimension of the Hilbert space is randomly taken from a Poisson distribution.
In other words, a given theory in the ensemble has the $N$ by $N$ zero matrix as its Hamiltonian, and the ensemble is formed by choosing $N$ from a Poisson distribution.
The only parameter of the theory is the mean $\lambda$ of the Poisson distribution from which we choose $N$.

This model is simple enough that we can calculate averages of some quantities of interest.
The $c_q$ are Schur polynomials in $N$ variables all taking the value 1.
In particular the elementary polynomial $c_{(1^n)}=e_n$ of degree $n$ is simply $\binom{N}{n}$.
Using $Z c_{(1^{n-1})} = c_{(2,1^{n-2})} + c_{(1^n)}$ (the relation \eqref{Schurrelation}) we get
\begin{equation}
	c_{(2,1^{n-2})} = Z e_{n-1} - e_{n} = N\binom{N}{n-1} - \binom{N}{n}
\end{equation}
In particular the ratios $c_{(2,1^{n-2})}/c_{(1^n)}$ are 
\begin{equation}
	\frac{c_{(2,1^{n-2})}}{c_{(1^n)}}
	=
	N\frac{n!(N-n)!}{(n-1)!(N-n+1)!} - 1
	=
	\frac{Nn}{N-n+1} - 1
	=
	\frac{(N+1)(n-1)}{N-n+1}
\end{equation}
Plugging this in to \eqref{fidelitybound}, we see the fidelity with which we can imitate a TFD state by states not connected between Alice and Bob is bounded by
\begin{equation}
	\begin{aligned}
		F_{\text{max}}
		&\geq
		1 - 
		\frac{n^2}{1 + (n-1)^2 + \frac{(N+1)(n-1)}{N-n+1} + (n-1)^2\frac{N-n+1}{(N+1)(n-1)}}
		\\
		&=
		\frac{n-1}{N(N-n-2)}.
	\end{aligned}
\end{equation}
This expression is valid for $n\leq N$.
The case $n>N$ can be obtained by a limit by considering Schur polynomials in $n$ variables $N$ of which take the value 1 and $n-N$ of which take a value $\epsilon$ which we take to zero.
The result is a lower bound of 1, so for $n>N$ we obtain $F=1$.
In other words, in a topological quantum mechanics theory with Hilbert space of dimension $N$, Alice and Bob can meet using superpositions with $N+1$ black holes.
In all we get the bound
\begin{equation}
	F_{\text{max}}
	\geq
	\begin{cases}
		\frac{n-1}{N(N-n-2)} & N\geq n\\
		1 & N<n
	\end{cases}
\end{equation}
The Poisson average of this expression,
\begin{equation}
	\expval{F_{\text{max}}}
	\geq
	e^{-\lambda}
	\left(
	\sum_{N=0}^{n-1}
	\frac{\lambda^N}{N!}
	+
	\sum_{N=n}^\infty
	\frac{\lambda^N}{N!}
	\frac{n-1}{N(N-n+2)}
	\right),
\end{equation}
where $\lambda$ is the mean of the Poisson distribution,
has a complicated closed form.
We can obtain a simpler bound by dropping the $N\geq n$ terms
\begin{equation}
	\expval{F_{\text{max}}}
	>
	\sum_{N=0}^{n-1}
	e^{-\lambda}
	\frac{\lambda^N}{N!}
	=
	\frac{\Gamma(n,\lambda)}{\Gamma(n)},
\end{equation}
where $\Gamma(n,\lambda)$ is the incomplete gamma function.
This partial sum is close to 1 when $n$ is significantly larger than the Poisson mean $\lambda$.
In fact, not only is the average fidelity bounded close to 1, the probability that the boundary theory has Hilbert space dimension less than $n$, and hence allows fidelity equal to 1, becomes high.
To be more precise, the function $\Gamma(n,\lambda)/\Gamma(n)$ of $n$ starts at 0 then grows to 1 as $n$ passes $\lambda$.
\begin{figure}[t]
	\centering
	\includegraphics[width=0.5\linewidth]{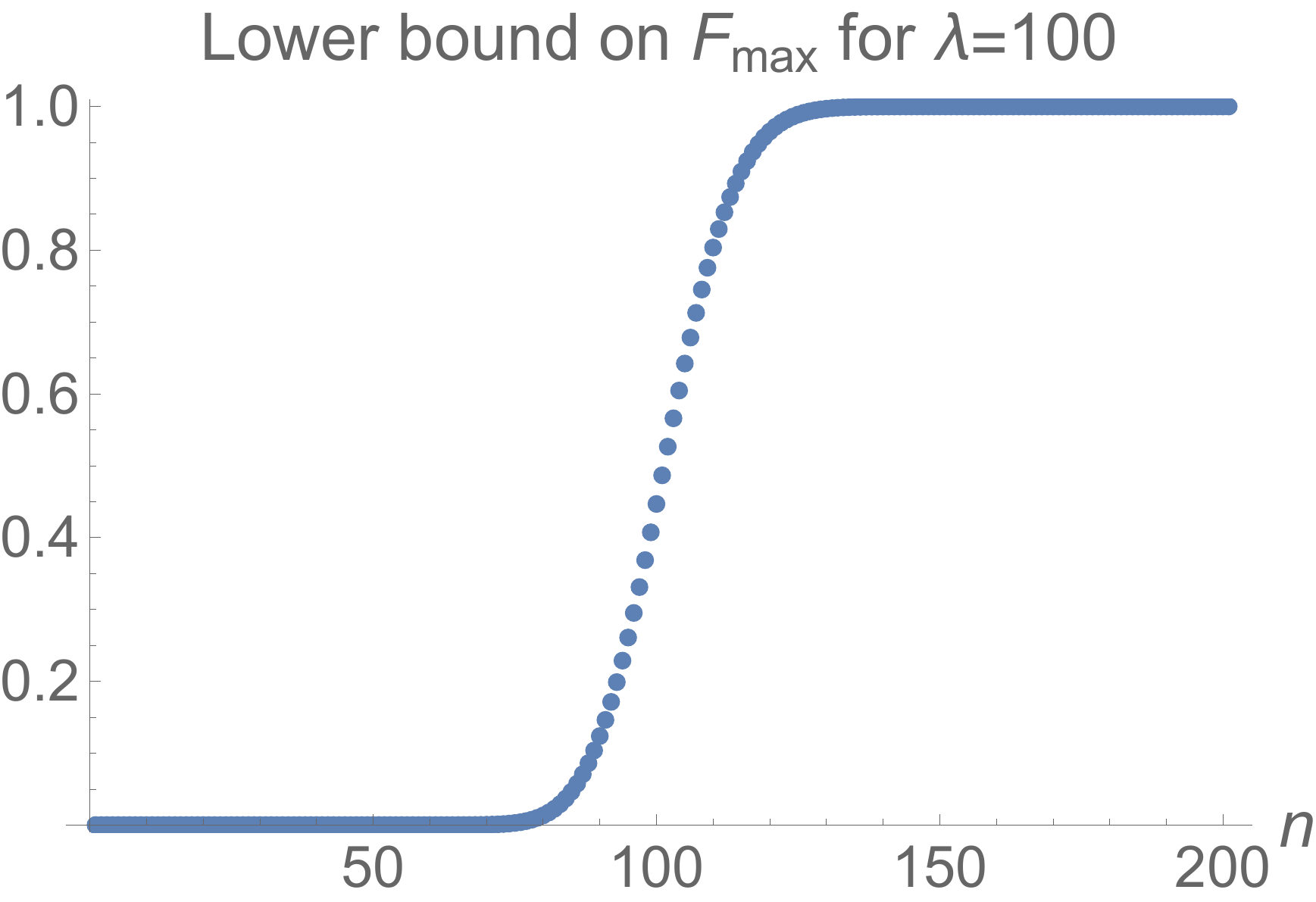}
	\caption{Lower bound on the maximum fidelity as a function of $n$, for the case $\lambda=100$ in the Marolf-Maxfield model.}
	\label{fig:MM_F_versus_n}
\end{figure}
The case $\lambda=100$ is plotted in figure \ref{fig:MM_F_versus_n}

The path integral in the Marolf-Maxfield model is a weighted sum over 2d spacetime topologies.
The model has a single parameter $S_0$, which determines the factor $e^{-2S_0}$ by which we weight each additional handle on a spacetime.
This is related to the mean dimension $\lambda$ by $\lambda=e^{2S_0}/(1-e^{-2S_0})$.
The name $S_0$ was chosen by analogy with JT gravity where the same weighting by number of handles occurs.
JT gravity can be understood as a dimensional reduction of gravitational dynamics of near-extremal black holes in higher dimensions \cite{Maldacena:2016upp, Almheiri:2014cka}.
In this context $S_0$ has the interpretation as the area of the black hole horizon and thus the entropy of the black hole.
The $S_0$ parameter is thus the conceptual stand-in for the black hole entropy in the much simpler Marolf-Maxfield model.
The $S_0$ parameter also has meaning in the matrix integral dual of JT gravity, where the $e^{S_0}$ controls the density of energy levels \cite{saad2019jt}.
In this sense the dependence on $S_0$ that we see is consistent with the general principle that closer energy spacing and higher temperature make achieving fidelity close to 1 harder.

\section{Random Hamiltonians}\label{randomhamiltonians}
As our third example we consider a system with random Hamiltonian.
Much work over the past few years has explored bulk gravity theories dual to statistical \emph{ensembles} of boundary theories.
For example, JT has been shown to be dual to a double-scaled random matrix integral \cite{saad2019jt}.
Here, however, we simply consider a theory with Hamiltonian taken from a Gaussian unitary ensemble, the goal being merely to ascertain typical behavior of the maximum fidelity \eqref{TFDfidelity}.

\begin{figure}[t]	
	\centering
	\begin{subfigure}[t]{0.45\linewidth}
		\centering
		\includegraphics[width=\linewidth]{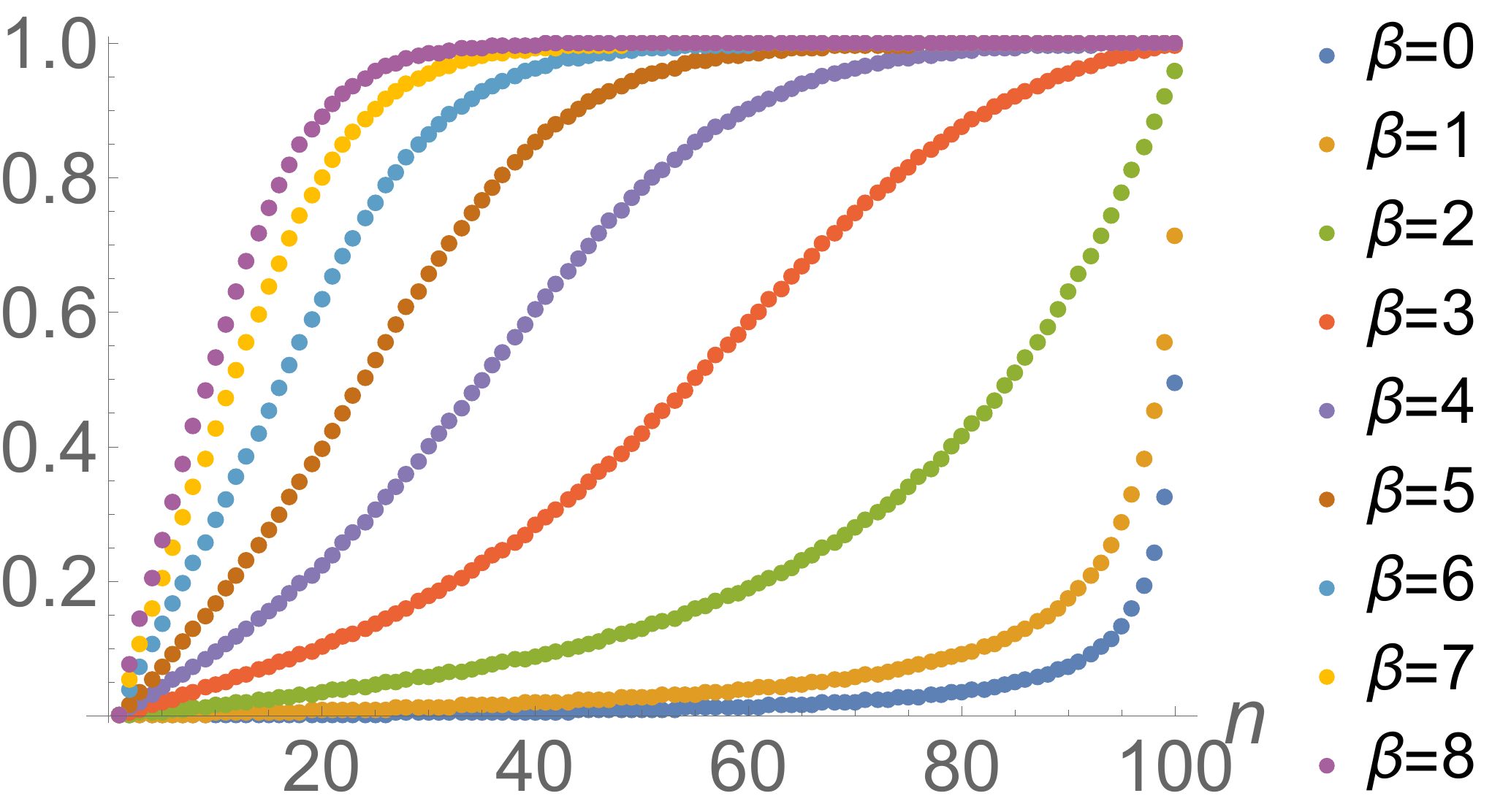}
		\subcaption{Bound on $F_\text{max}$ as a function of $n$ for various $\beta$, with $N=100$}
		\label{fig:rand_N_100}		
	\end{subfigure}
	\quad
	\begin{subfigure}[t]{0.45\linewidth}
		\centering
		\includegraphics[width=\linewidth]{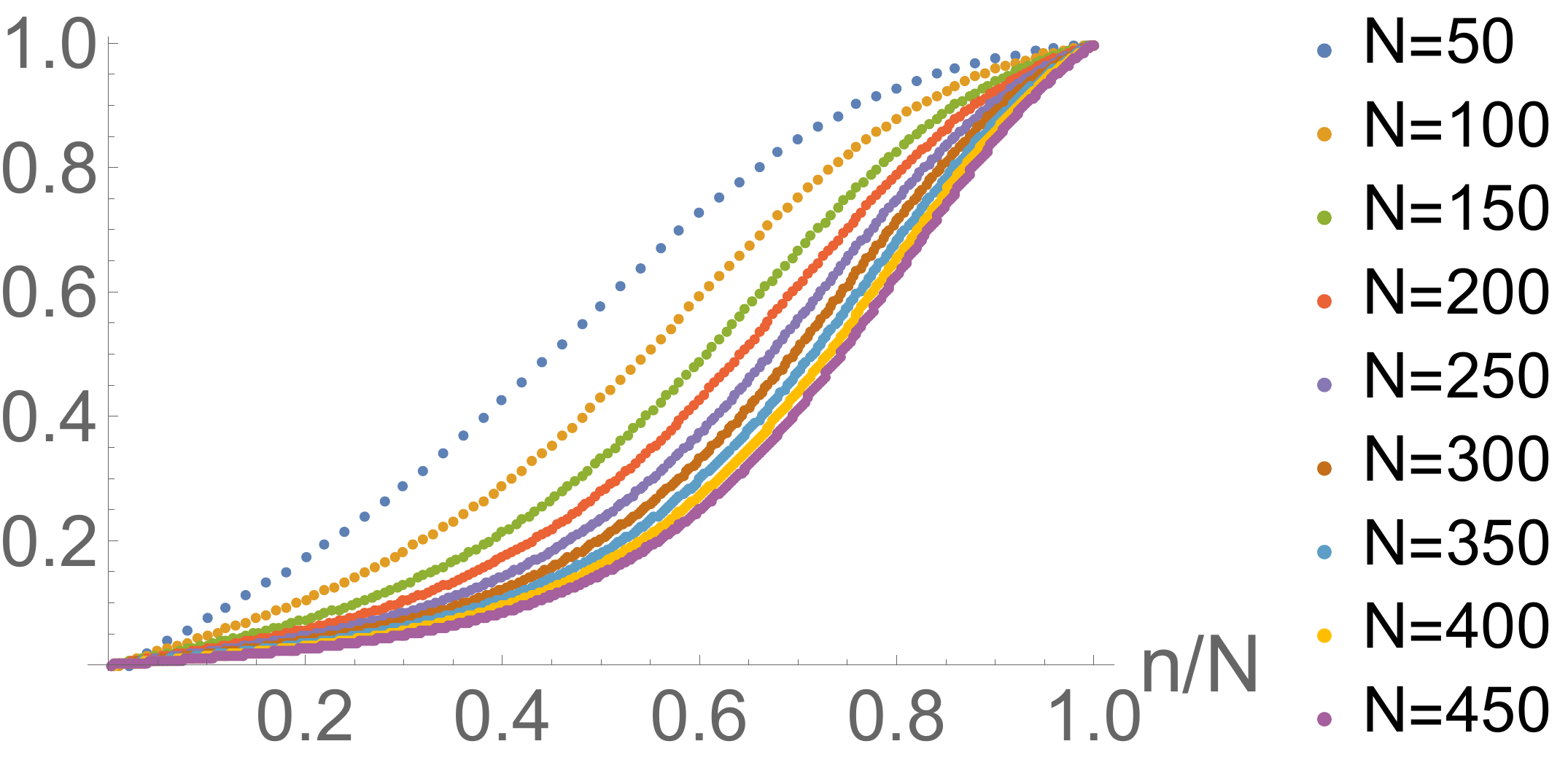}
		\subcaption{Bound on $F_\text{max}$ as a function of $n/N$ for various $N$, with $\beta=3$}
		\label{fig:rand_beta_3}
	\end{subfigure}
\end{figure}

We numerically calculated the bound \eqref{fidelitybound} for various choices of $\beta$ and Hilbert space dimension $N$.
Elementary symmetric polynomials were calculated numerically using the recursion relation described in Section \ref{harmonicoscillator}.
(In fact, other desired Schur polynomials can be calculated accurately and efficiently from the elementary symmetric polynomials using the dual Jacobi-Trudi formula \cite{demmel2006accurate}.)
Some results are plotted in figures \ref{fig:rand_N_100} and \ref{fig:rand_beta_3}.
The $n$ required for $F_\text{max}$ to be bounded near 1 depends on both $\beta$ and $N$.
For small $\beta$, $n$ must be nearly $N$.
For a given $\beta$, the required $n$ as a fraction of $N$ increases with $N$, but seems to converge as $N$ gets large.

\section{Black holes with $T_{HP}<T<2T_{HP}$}\label{semiclassical}
Now take as our system a holographic CFT on the spatial sphere $S^{d-1}$.
If we consider just the semiclassical approximation of the gravity contributions to the path integral, for low temperatures the partition function will be dominated by the saddlepoint corresponding to thermal AdS while for high enough temperatures the dominant saddlepoint will be a Euclidean AdS-Schwarzschild black hole \cite{Hawking:1982dh, Witten:1998zw}.
Taking just the background contributions of these saddlepoints we get an approximation for the partition function
\begin{equation}
	Z(\beta)
	\approx
	\begin{cases}
		e^{-I[\text{Schwarzschild}]} & \beta < 1/T_{HP}\\
		e^{-I[\text{AdS}]} & \beta > 1/T_{HP}
	\end{cases},
\end{equation}
where $I$ is the classical gravitational action of the background and $T_{HP}$ is the Hawking-Page transition temperature.

Recall that the inner products $\expval{\pi}$ equal products of factors of the form $Z(k\beta)$ where $k$ is a positive integer.
Given a fixed $\beta_0$, the coefficients $c_q$ thus only depend on the partition function evaluated at the points $k\beta_0$.
Choose $\beta_0$ such that $\beta_0<1/T_{HP}$ and $2\beta_0>1/T_{HP}$.
Note that $I[\text{AdS}]$ is of the form $e^{c\beta}$ where $c$ is constant in $\beta$ \cite{Witten:1998zw}.
Rescaling $Z(\beta)$ by a factor $e^{c\beta}$ does not change the maximum fidelities; it is merely a change in normalization corresponding shifting the energies by a constant $c$.
After fixing $\beta_0$ and rescaling we get the form
\begin{equation}\label{Zform}
	Z(k\beta_0)
	\approx
	\begin{cases}
		z & k = 1\\
		1 & k > 1, k\in \N
	\end{cases},
\end{equation}
where $z=\exp(I[\text{AdS}]-I[\text{Schwarzschild}])|_{\beta=\beta_0}$ is a value greater than 1 that depends on the choice of $\beta_0$.
This form for the values $Z(k\beta_0)$ simplifies the expressions for the inner products and allows us to calculate the coefficients $c_{(1^n)}$ and $c_{(n)}$ corresponding to the sign and trivial irreps respectively.
From the expressions \eqref{expvaldefinition} for the inner products, it is straightforward to calculate generating functions
\begin{equation}
	\sum_{n=0}^\infty c_{(1^n)}t^n
	=
	(1+t)e^{(z-1)t}
\end{equation}
and
\begin{equation}
	\sum_{n=0}^\infty c_{(n)}t^n
	=
	\frac{1}{1-t}e^{(z-1)t},
\end{equation}
from which we extract the coefficients $c_{(1^n)}=(z-1)^{n-1}(z-1+n)/n!$ and $c_{(n)}=\sum_{m=0}^n(z-1)^m/m!$.
Using these values for $c_{(1^n)}$ in \eqref{fidelitybound} gives a bound on fidelity that decreases as $n$ gets large.
The true maximum fidelity cannot, of course, decrease as $n$ increases.
This signifies either a deficiency in the bound \eqref{fidelitybound} or a deficiency in our approximate form \eqref{Zform}.
Using the above values for $c_{(n)}$ in \eqref{fidelityboundmaxp} also fails to give a bound that increases to $1$;
In this case, as $n$ increases the bound approaches $1-\frac{1}{z}$.
Again, we either require a better bound on fidelity or a better estimate for the values $Z(k\beta_0)$ in order to see the fidelity increase to 1.
We will see a bound that accomplishes this in the next section.

\section{Bounds on fidelity from subgroups}\label{subgroup}
Using expression \eqref{nearlynull} for the fidelity of writing one geometry as a superposition of the others requires knowing the coefficients $c_q$ for the given system and temperature.
This is in general difficult.
In particular, considering limits in large $n$ requires knowing the $c_q$ for the groups $S_n$ for arbitrary $n$, not likely to be simple in general.
One option to proceed is to consider a subgroup of $S_n$ in the hopes that the coefficients $c_q$ for irreps of the subgroup are easier to find.
Let $G$ be a subgroup of $S_n$, and let $F_G$ be the maximum fidelity between $\ket{\mathrm{id}}$ and a superposition of states $\{\ket{\pi}|\pi\in G, \pi\neq \mathrm{id}\}$.
Then $F_G\leq F_{\text{max}}$, as we are maximizing over a smaller subspace of states.
Following the procedure in Section \ref{nearlynullstates}, but for an arbitrary subgroup $G$ of $S_n$, we find
\begin{equation}\label{nearlynullG}
	F_G=1-\frac{\left|G\right|^2}{Z^n\sum_q d_q^3/c_q},
\end{equation}
where the sum is over irreps $q$ of $G$.

We'd like to choose a subgroup $G$ that is large enough to potentially give an approximate null state, but that has an easily understandable representation structure.
Consider the choice
\begin{equation}
	G=\big\{(1\ 2)^{a_1}(3\ 4)^{a_2}\cdots (n\!-\!1\ n)^{a_{n/2}}\mid a_i\in\{0,1\}\big\}.
\end{equation}
For ease of notation we can refer to an element of $G$ by the $(n/2)$-tuple $\vec{b}=\left(b_1,b_2,\ldots,b_{n/2}\right)$ with entries either $0$ or $1$.
The group $G$ is abelian so its irreps are very simple.
Like elements of $G$ they are labeled by $(n/2)$-tuples with entries either $0$ or $1$, notated $\vec{\gamma}=\left(\gamma_1,\gamma_2,\ldots,\gamma_{n/2}\right)$.
They are given by
\begin{equation}
	\chi_{\vec{\gamma}}(\vec{b})
	=
	\chi_{\left(\gamma_1,\ldots,\gamma_{n/2}\right)}(b_1,\ldots,b_{n/2})
	=
	\left(-1\right)^{\sum_i\gamma_i b_i}.
\end{equation}
The coefficients $c_{\vec{\gamma}}$ are
\begin{equation}
	\begin{aligned}
		c_{\vec{\gamma}}
		&=
		\frac{1}{2^{n/2}}\sum_{\vec{b}}\vec{\gamma}(\vec{b})\big<\vec{b}\big>
		=
		\frac{1}{2^{n/2}}\sum_{\vec{b}}
		(-1)^{\sum_i\gamma_i b_i}
		Z(\beta)^n
		\prod_{i=1}^{n/2}
		\left(Z(2\beta)/Z(\beta)^2\right)^{b_i}\\
		&=
		\frac{Z^n}{2^{n/2}}
		\prod_{i=1}^{n/2}
		\left(
		1
		+
		(-1)^{\gamma_i}
		\frac{Z(2\beta)}{Z(\beta)^2}
		\right).
	\end{aligned}
\end{equation}
As $G$ is abelian the dimension of every irrep is 1.
Applying \eqref{nearlynullG} results in
\begin{equation}
	\begin{aligned}
		F_G
		=
		1
		-
		\left(
		1
		-
		\frac{Z(2\beta)^2}{Z(\beta)^4}
		\right)^{n/2}
	\end{aligned}
\end{equation}
Note that $Z(2\beta)/Z(\beta)^2$ is less than 1.
So as $n$ gets large $F_G$, and hence $F_\text{max}$ approaches 1.

Given a value of $\beta$, we can ensure $F_G>1-\epsilon$ and hence $F_\text{max}>1-\epsilon$, for any small $\epsilon>0$ by choosing
\begin{equation}\label{sufficient}
	n>\frac{2\log\epsilon}{\log(1-\frac{Z(2\beta)^2}{Z(\beta)^4})}.
\end{equation}

\subsection{Dependence on Newton's constant $G_N$}
In cases where the partition function takes the form $Z(\beta)\sim e^{f(\beta)/G_N+\mathcal{O}(G_N)}$, the requirement \eqref{sufficient} implies that $\log n$ must grow like $1/G_N$ for small $G_N$.
Achieving the fidelity bound $F_G$ thus requires a superposition with $e^{e^{\mathcal{O}(1/G_N)}}$ terms in it, so doubly exponential in $1/G_N$.
That an approximate null state would require the superposition of a large number of states is consistent with the results of \cite{Almheiri:2016blp}.
They found that the entropy is linear to leading order on superpositions of much fewer than $e^{\mathcal{O}(1/G_N)}$ geometric states, and that this linearity can break down for superpositions of on order $e^{\mathcal{O}(1/G_N)}$ states.
Our null state resulting in fidelity $F_G$, on the other hand, requires a much larger superposition.

This suggests the possibility that a much better bound is achievable.
In the Marolf-Maxfield model, for example, $Z(2\beta)/Z(\beta)^2$ is $1/N$ where $N$ is Poisson random with mean $\lambda$.
This leads to a bound $F_G(N) = 1 - (1-1/N)^{n/2}$, whereas we know that the true maximum fidelity is 1 as soon as $n>N$.
Thus $F_\text{max}$ will be close to 1 when $n$ is significantly greater than the average dimension $\lambda$.
Taking $S_0\sim 1/G_N$ then suggests a superposition with merely $e^{\mathcal{O}(\frac{1}{G_N}\log\frac{1}{G_N})}$ terms.

\section{Additional discussion}
As explained in the introduction, the possibility of rewriting the TFD state as a partial trace of a superposition $\sum_{\pi(1)\neq 1}\alpha_\pi\ket{\pi}$ leads to a situation where Alice and Bob could meet despite the fact that they do not meet in any term of the superposition.
Can a superposition of worlds where Alice does not meet Bob really equal a world where she does?
This is of course possible if there is not in fact a ``meeting" quantum observable.
This in turn is reasonable if geometry is itself not a quantum observable.

One thing to point out is that, while the fidelity between two states being close to 1 bounds correlators of the states to be close to each other, this bound on correlators is not uniform.
That is to say, depending on the quantity one desires to measure, one may need a larger or smaller number $n$ to get outcomes within a desired error.
It's possible that a putative ``meeting" operator might have an especially stringent requirement on $n$.

The question of what Alice observes must start with an identification of ``Alice" within the system.
It's possible that there is more than one way to do this or that this identification isn't linear.
Perhaps the concept of an observer ``Alice" is state-dependent.
This might neatly solve the paradox.
The paradox arises from the fact that Alice, by recording whether or not she meets Bob, seems to be recording a fact about the entanglement of the state.
For an observer as usually understood, this cannot be the case, as entanglement is not linear on the space of states.
But perhaps there's nothing wrong with a state-dependent sense of ``Alice" being able to measure a nonlinear observable.

Throughout this work we have assumed the duality between two-sided eternal black holes and TFD states.
Two TFD states with different temperatures will have a nonzero inner product even though they correspond to different geometries: two black holes with different horizon area.
If there is not a geometry quantum observable, this is not troubling.
If, however, we are set on the existence of a quantum operator for geometry, we can make sense of the nonzero overlap by interpreting the TFD state as dual to a wave function over different, orthogonal (by supposition) geometries where the peak amplitude is at the black hole geometry with the appropriate temperature.
Then the nonzero inner product measures fluctuations away from the peak geometry, and the smallness of the overlap
\begin{equation}
	\frac{|\braket{\text{TFD}(\beta_1)}{\text{TFD}(\beta_2)}|^2}{\braket{\text{TFD}(\beta_1)}\braket{\text{TFD}(\beta_2)}}
	=
	\frac{Z\big((\beta_1+\beta_2)/2\big)^2}{Z(\beta_1)Z(\beta_2)}
\end{equation}
when $\beta_1$ and $\beta_2$ are very different simply signifies that in the wavefunction dual to $\ket{\text{TFD}(\beta_1)}$ the amplitude of the black hole geometry with $\beta_2$ is small (and vice versa).
This point of view is less able to explain the nonzero overlaps between the $\ket{\pi}$ states, however.
By locality, the only geometries that should appear in the wave function dual to $\ket{\text{TFD}}\otimes\ket{\text{TFD}}$ should be geometries made of two disjoint components each of which appears in the wavefunction for $\ket{\text{TFD}}$. 
In particular, no geometries connecting the boundary of the first tensor factor with the boundary of the second should ever appear.
But the wave function for $(1\ 2)\circ\ket{\text{TFD}}\otimes\ket{\text{TFD}}$, two copies of the TFD state with their right sides swapped, will \emph{only} have support over such geometries, again, by locality of the boundary theory.
Why then the nonzero overlap $\braket{\mathrm{id}}{(1\ 2)}=Z(2\beta)/Z(\beta)^2$?
If we are to understand these states as wavefunctions over orthogonal states labeled by geometries then the wavefunction of $\ket{\text{TFD}}\otimes\ket{\text{TFD}}$ is not simply two copies of the wavefunction for $\ket{\text{TFD}}$.\footnotemark
\footnotetext{
	This is reminiscent of the factorization problem that generically appears in gravity path integrals that sum over geometries \cite{Maldacena:2004rf}, though it is not quite the same, as the nonzero overlap $\braket{\mathrm{id}}{(1\ 2)}$ in question has a single boundary spacetime component.
	Instead, this is a nonfactorization at the level of Hilbert spaces (codimension 1) rather than at the level of partition functions (codimension 0).
}
One natural conclusion is that there simply is no geometry operator and corresponding complete set of states labeled by geometry.
Another natural conclusion is that different geometry states are related by null states, which amounts to the same thing.

The rewriting of a thermofield double state described in this work may be interesting in the context of traversable wormholes as described in \cite{Gao:2016bin} and \cite{Maldacena:2017axo}.
In these cases information can travel from one side to the other via an interaction that is introduced between the boundaries.
From the bulk point of view, this interaction changes the geometry to include a shockwave that makes the wormhole traversable.
This bulk point of view no longer holds once we have rewritten the TFD state to be a superposition of states where the two sides are not connected.
Is there a bulk explanation as to how information can traverse that makes sense within each term of the superposition?

\bibliographystyle{JHEP}
\bibliography{nullstates}

\end{document}